\documentclass[useAMS,usenatbib]{mnras}

\usepackage{graphicx}   
\usepackage{amsmath}    
\usepackage{amssymb}    
\usepackage{multicol}   
\usepackage{bm}         
\usepackage{pdflscape}  
\usepackage{float}      
\usepackage{tipa}       
\hypersetup{final}

\title[]
{Sub-mm free-free emission from the winds of massive stars in the age of ALMA}
\author[]{Daley-Yates S., Stevens I. R., Crossland T. D.}

\author[S. Daley-Yates, I. R. Stevens and T. D. Crossland]{S. Daley-Yates$^{1}$\thanks{E-mail:
sdaley@star.sr.bham.ac.uk}, I. R. Stevens$^{1}$ and T. D. Crossland$^{1}$\\
$^{1}$School of Physics and Astronomy, University of Birmingham, Edgbaston, Birmingham, B15 2TT\\}

\begin{document}

\date{}

\pagerange{\pageref{firstpage}--\pageref{lastpage}} \pubyear{2015}

\maketitle

\label{firstpage}

\begin{abstract}
The thermal radio and sub-mm emission from the winds of massive stars is investigated and the contribution to the emission due to the stellar wind acceleration region and clumping of the wind is quantified. Building upon established theory, a method for calculating the thermal radio and sub-mm emission using results for a line-driven stellar outflow according to \citet{CAK1975} is presented. The results show strong variation of the spectral index for $10^{2}$~GHz $< \nu <$ $10^{4}$~GHz. This corresponds both to the wind acceleration region and clumping of the wind, leading to a strong dependence on the wind velocity law and clumping parameters. The Atacama Large Millimeter/sub-mm Array (ALMA) is the first observatory to have both the spectral window and sensitivity to observe at the high frequencies required to probe the acceleration regions of massive stars. The deviations in the predicted flux levels as a result of the inclusion of the wind acceleration region and clumping are sufficient to be detected by ALMA, through deviations in the spectral index in different portions of the radio/sub-mm spectra of massive stars, for a range of reasonable mass-loss rates and distances. Consequently both mechanisms need to be included to fully understand the mass-loss rates of massive stars.
\end{abstract}


\section{Introduction} 
\label{sec:intro}

The stellar winds of massive stars are driven primarily by absorption in spectral lines and are therefore known as line-driven winds. The theoretical basis 
for the description of these winds was set out by \citet[hereafter CAK]{CAK1975}, building on the work by \cite{Lucy1970}. These winds are distinctive due to a non-linear dependence of the wind acceleration on the local density and velocity gradient. As a result, the wind structure is highly unstable and dynamic in nature \citep{Runacres2002}. These effects make diagnosing the wind structure challenging and their theoretical treatment is an active area of research.

Observations of massive stellar winds are a principle means for diagnosing their properties \citep{Barlow1977, Prinja1988, Castor1979}. Both thermal and non-thermal emission is detectable from these stellar winds, but thermal emission provides a large spectral window for characterising the wind properties \citep{Wright1975,De Becker2007}. Radio emission from massive stars has historically been the subject of considerable interest, \citep{Braes1972,Wright1974,Cohen1975} and in particular \citet[hereafter WB75]{Wright1975}. Thermal emission from massive stars is not due to processes at the optical photosphere, but free-free interactions between charged species in the ionised wind material \citep{De Becker2007}. Therefore, any predictions of thermal emission assumes that the stellar surface is radio quiet. This work will explore the region where radio and sub-mm emission is due to multiple factors including the stellar blackbody spectrum and wind acceleration region. 

Analytical modelling of the symbiotic nova V1016 Cyg was accomplished by \cite{Seaquist1973}. This work provided the first early steps towards characterising the radio thermal spectral flux from stellar objects. \cite{Seaquist1973} assumed a uniform, spherically symmetric, time-independent, isothermal flow. The resulting spectral flux density as a function of frequency takes the form $S_{\nu} \propto \nu^{\alpha}$, where $-0.1 \leq \alpha \leq +2$. This model was built upon in the seminal work conducted by WB75. This highly successful model for the prediction of thermal emission from stellar winds is frequently quoted to explain observational results and justify theoretical conclusions \citep{Blomme2003, Leitherer1991, Montes2011}. Refinement of the \cite{Seaquist1973} model by WB75 leads to a spectral index $\alpha = 0.6$. 

WB75 describe how their model agrees well with observation of a selection of massive stars at GHz radio frequencies. Their model is based upon the assumption that the flux originates from the outer reaches of the wind (but still sufficiently close to the star to be considered isothermal) where the velocity can be approximated at all radii, by the wind terminal velocity. Beyond this region, the temperature of the wind is insufficient for full ionisation of the wind and recombination occurs \citep{Drew1989}. At this point, the free-free interactions cease and thermal emission is extinguished.

As the observational frequency increases, emission from the accelerating wind begins to contribute to $S_{\nu}$. WB75 discuss this and state that an analytic solution which accounts for the acceleration region is not possible. In their paper, WB75 compare the model predictions to both radio and infrared observations of the star P Cygni. They find that the slopes of both data sets independently agree with their predictions. However, it is not until the model is extrapolated from low to high frequencies, that there is a higher level of flux at infrared wavelengths than the model predicts. This implies a steepening in the spectral index.

Several observations conducted at frequencies up to 250~GHz for number of different stellar objects have been carried out. For example, observations of 
Wolf-Rayet binaries by \cite{Montes2015}, however, the focus of the study is the wind-wind collision region and not the initialisation of the wind. Observations of single massive stars conducted at 230~GHz were carried out by \cite{Leitherer1991}. Acceleration and deceleration regions are considered but in the context of the extended wind and not the initialisation region. One study at sub-mm and infrared frequencies of the Wolf-Rayet star $\gamma$ Velorum by \cite{Williams1990} shows clear deviation from $\alpha = 0.6$ at high frequencies. \cite{Nugis1998} see a steepening in the spectra with $\alpha = 0.77$ and $0.75$ for the winds of a sample of WN and WC stars respectively, with the deviation from $\alpha = 0.6$ attributed to clumping of the wind material. Their data is very sparse in the sub-mm range however.

A number of infrared observations of early type stars have been conducted by both \cite{Castor1983} and \cite{Abbott1984}. \cite{Castor1983} deduced that their observations could not determine the stellar wind velocity laws, leading to large uncertainties in the mass-loss rates of the stars they studied. \cite{Abbott1984} found that the velocity law varies dramatically from star to star, again leading to the large variations in the mass-loss rates. The above examples highlight the importance of the acceleration region as an avenue of exploration into the properties of early type stars and the importance of future observations.

\cite{Pittard2010} briefly discuss the presence of the acceleration region and its impact on the spectral flux, as a preamble to a study of binary O stars. Consequently the calculations are limited to a single set of stellar parameters. However, the frequency range covered in this calculation is extensive and includes the acceleration region.

Telescope technology as been the main limitation to this exploration. With the advent the Atacama large Millimeter/sub-mm Array (ALMA), this situation has changed. ALMA has both simultaneously the spectral window and sensitivity to allow for observations of massive stars which test the full spectrum of their wind. ALMA has several bands covering the range from 84~GHz to 950~GHz. As an example of ALMAs sensitivity, at an observing frequency of 630 GHz (Band 9), with a bandwidth of 7.5 GHz and dual polarisation, a rms sensitivity of $\sim 0.25$~mJy is achieved with an integration time of 3600 s. These calculations were performed assuming optimal observing conditions, for example lowest water vapor column density and were conducted using the ALMA Sensitivity Calculator available on the ALMA website.

Presented in this work are numerical calculations of the thermal free-free emission from the stellar winds of an ensemble of massive stars. Both accelerating and terminal velocity regions of the wind are accounted for together with a consideration of non-smooth clumped winds. The results are placed into the context of what is observable by ALMA. Comparisons are drawn between the numerical results and the WB75 model. The following section will give a brief overview of the stellar parameters which have been used in this study.

\section{Stellar Parameters}

\begin{figure}
\centering
\includegraphics[width=0.5\textwidth]{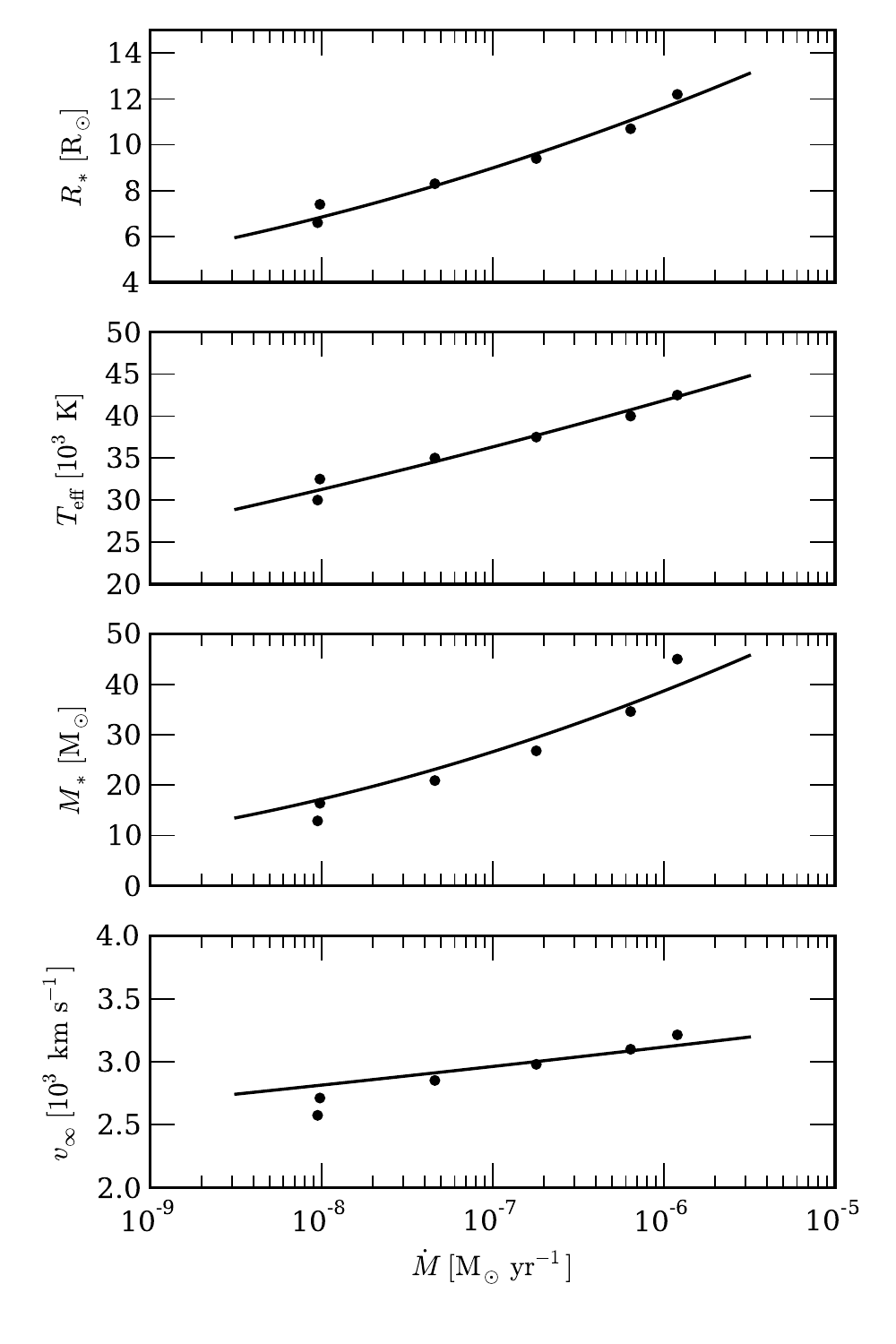}
\caption{
Stellar parameters used by both the WB75 analytic model and numerical model presented in this work, to calculate the spectral flux density for a series of massive stars with mass-loss rate in the range $10^{-8.5}$~M$_{\sun}$~yr$^{-1}$ $< \dot{M} <$ $10^{-5.5}$~M$_{\sun}$~yr$^{-1}$. The data are taken from \protect\cite{Krticka2012}.
\label{fig:stellar_parameters}
}
\end{figure}

The following stellar parameters are taken from \cite{Krticka2012} in the case of the O-type stars and from \cite{Krticka2014} in the case of the B-type stars. The values of these parameters are plotted as a function of mass-loss rate, $\dot{M}$, in Fig. \ref{fig:stellar_parameters}. The stellar parameters of the B-type stars are not displayed as, while they were used to place constraints on the process described below, they are outside the range of stellar parameters used in the calculations. \cite{Krticka2014} and \cite{Krticka2012} derive these stellar parameters from a model non-local thermodynamic equilibrium (NLTE) stellar wind, which they use to derive values of $\dot{M}$. Their values for the stellar effective temperature, $T_{\mathrm{eff}}$, stellar radius, $R_{*}$ and stellar mass, $M_{\ast}$, are interpolated from formulas derived by \cite{Harmanec1988}.

Values for the terminal velocity are arrived at by assuming that $v_{\infty} \approx 3 v_{\mathrm{esc}}$, where $v_{\mathrm{esc}}$ is the stellar escape velocity, with $v_{\mathrm{esc}} = \sqrt{2 G M_{\ast} (1 - \Gamma_{\mathrm{e}})/R_{\ast}}$. $\Gamma_{\mathrm{e}}$ is the Eddington parameter of the star which is derived in turn from the stellar luminosity; $\Gamma_{\mathrm{e}} = \sigma_{\mathrm{e}} L_{\ast} /(4 \pi \mathrm{G} M_{\ast}) $ where it is assumed that $L_{\ast} = 4 \pi R_{\ast}^{2} \sigma T_{\mathrm{eff}}^{4}$ with $\sigma$ the Stefan-Boltzmann constant. $\sigma_{\mathrm{e}}$ is the electron scattering opacity and $G$ is the gravitational constant. Throughout this study the distance between the star and the observer will be kept constant for all stellar models at $D = 0.5$~kpc.

By plotting each of these stellar parameters as a function of $\dot{M}$, a series of polynomials can be optimised via the least squares method to give a functional relationship between $\dot{M}$ and $T_{\mathrm{eff}}$, $R_{\ast}$ and $M_{\ast}$. The polynomial fitted is 2\textsuperscript{nd} order in $\log_{10}(\dot{M})$ and takes the form
\begin{equation}
	f(\dot{M}) = a + b \log_{10}(\dot{M}) + c \left( \log_{10}(\dot{M}) \right)^{2},
	\label{eq:m_dot_fit}
\end{equation}
where $f(\dot{M})$ is the stellar parameter that is being calculated and $a$, $b$ and $c$ are fit parameters which undergo optimisation. Equation (\ref{eq:m_dot_fit}) is also shown in Fig. \ref{fig:stellar_parameters} as the black line in each graph.

Equation (\ref{eq:m_dot_fit}) assumes an oversimplified relationship between $\dot{M}$ and the stellar parameters. For example, the model only accounts for main sequence B-type and O-type stars and not Giants or Supergiants. However, it has captured the essential scaling between $\dot{M}$ and the other parameters over the range of $\dot{M}$ values considered in this work.

Some care is required when considering the range of $\dot{M}$ values to use during the calculations. A star with a high mass and correspondingly high $\dot{M}$ together with an intense luminosity, will have a large effective radius at radio wavelengths (which will be discussed in Section \ref{sec:radio_emission_symmetry}) even at high radio frequencies and the opposite is true for a low mass star. As such, the influence of the acceleration region, may not be within the observable frequencies of the current generation of radio telescopes. To account for this, a suitable range for $\dot{M}$ was chosen and found to be $10^{-8}$ M$_{\sun}$ yr$^{-1}$ $< \dot{M} <$ $10^{-5.5}$ M$_{\sun}$ yr$^{-1}$, with six models designated S0~-~S5, where "S" refers to "smooth", with mass-loss rates evenly spaced by 0.5 dex. Table \ref{tab:parameters} displays the final parameters that were used to perform the calculations set out below. For all models the mean ion charge $Z~=~1.128$ and the ratio of electron and ion number densities $\gamma~=~1.09$, were kept constant.

\begin{table*}
\caption{The assumed stellar parameters for the calculated models. \label{tab:parameters}}
\begin{tabular}{ccccccccc}
\hline
Model & $\dot{\mathrm{M}}$ $[\mathrm{M}_{\sun} / \mathrm{yr}]$ & $R_{\ast}$ $[R_{\sun}]$ & $T_{\mathrm{eff}}$ $[10^{3}$ $\mathrm{K}]$ & $M_{\ast}$ $[M_{\sun}]$ & $v_{\infty}$ $[10^{3}$ $\mathrm{km \ s^{-1}}]$ & $v_{\mathrm{esc}}$ $[10^{3}$ $\mathrm{km \ s^{-1}}]$ & $\log_{10}(L_{\ast}/L_{\sun})$ & $\Gamma_{\mathrm{e}}$ \\
\hline
S0 & $10^{-5.5}$ & 13.1 & 44.7 & 45.7 & 3.20 & 1.07 & 5.75 & 0.31 \\
S1 & $10^{-6.0}$ & 11.6 & 41.9 & 38.8 & 3.12 & 1.04 & 5.53 & 0.22 \\ 
S2 & $10^{-6.5}$ & 10.2 & 39.0 & 32.3 & 3.04 & 1.01 & 5.30 & 0.15 \\
S3 & $10^{-7.0}$ &  9.0 & 36.3 & 26.6 & 2.96 & 0.99 & 5.06 & 0.11 \\ 
S4 & $10^{-7.5}$ &  7.9 & 33.7 & 21.6 & 2.89 & 0.96 & 4.82 & 0.08 \\ 
S5 & $10^{-8.0}$ &  6.8 & 31.3 & 17.2 & 2.81 & 0.94 & 4.56 & 0.05 \\
\hline
\end{tabular}
\end{table*}

By investigating the above range of possible $\dot{M}$ values, it can be determined whether any deviation from a constant spectral index is observed when a non-terminal velocity flow is taken into account and within the range of current observatories.

\section{Emission Models} 

This section will review the theoretical basis of both terminal velocity and accelerating winds, followed by a discussion of the concept of a frequency dependent effective stellar radius. The Gaunt factor plays an important role in these concepts and as such a review of recent high precision calculation of this factor is given. The effect of clumping on the spectral flux is also considered and a limited set of calculations presented. Finally, the numerical setup of the calculations are communicated.

\subsection{Terminal Velocity Stellar Wind}
\label{sec:radio_emission_symmetry}

The model developed by WB75 assumes the winds has both spherical symmetry as well as terminal velocity. The principle results of this theory are outlined below.

The spectral flux density, $S_{\nu}$, at a distance $D$ from the star is given by the integral of the intensity of radiation, $I(\nu, T)$, along a line of sight from an observer to the star:
\begin{equation}
        \label{eq:flux_full}
        S_{\nu} = \int_{0}^{\infty} \frac{I(\nu, T)}{D^{2}} 2 \pi q \mathrm{d}q.
\end{equation}
Here the impact parameter $q$ gives the radial distance from the star, perpendicular to the line of sight of the observer in a cylindrical geometry. A rigorous treatment of the solution to this integral is given in WB75. For the purpose of this work it is sufficient to simply state the result:
\begin{equation}
        \label{eq:total_flux}
        S_{\nu} = 23.2 \left( \frac{ \dot{M} }{ \mu v_{\infty} } \right)^{4/3} 
        \frac{ \nu^{2/3} }{ D^{2} } \left( \gamma \mathrm{\textit{\textg}}_{\mathrm{ff}} \overline{Z^{2}} \right)^{2/3}\,.
\end{equation}
Here $\mu$ is the mean atomic weight of the gas, $v_{\infty}$ is the terminal velocity of the outflowing stellar material in $\mathrm{km}$~$\mathrm{s}^{-1}$, $\nu$ is the frequency of emitted radiation in $\mathrm{Hz}$, $D$ is the distance of the object from the observer in $\mathrm{kpc}$, \textit{\textg}$_{\mathrm{ff}}$ is the free-free Gaunt factor (see Section \ref{sec:General_radio}), $\gamma$ is the ratio of the electron and ion number densities, $\overline{Z^{2}}$ is the mean squared ion charge and the flux, $S_{\nu}$, is measured in Jy. Equation (\ref{eq:total_flux}) is valid in the region of the spectrum in which $h \nu \ll k_{\mathrm{B}} T_{\mathrm{eff}}$, limiting its applicability to infrared and radio frequencies. 
 
The above analysis leads to the spectral index $\alpha = 2/3$, and therefore $S_{\nu} \propto \nu^{2/3}$. However, the Gaunt factor, \textit{\textg}$_{\mathrm{ff}}$, has a slight frequency dependence (see Section \ref{sec:gaunt}). With this additional consideration $S_{\nu} \propto \nu^{0.6}$, at GHz frequencies.

Rearrangement of equation (\ref{eq:total_flux}) allows for the definition of an effective radius, $R_{\nu}$ (WB75), which represents the inner limit from which emission can propagate through the wind to the observer. Therefore, at a given frequency the total flux emitted is due to the material exterior to $R_{\nu}$. Equation (\ref{eq:effective_radius}) gives this radius in terms of the wind parameters discussed above:
\begin{equation}
        R_{\nu} = 2.8 \times 10^{28} \left( \gamma g \overline{Z^{2}} \right)^{1/3} T^{-1/2} \left( \frac{\dot{M}}{\mu v_{\infty} \nu} \right)^{2/3},
        \label{eq:effective_radius}
\end{equation}
where all the symbols have the same units as above and $R_{\nu}$ is measured in cm. As has already been mentioned, the model constructed by WB75 makes a number of assumptions about the geometry, composition and homogeneity of the circumstellar material, such as spherical symmetry and terminal wind velocity. See Section \ref{sec:effect_radius} for an in depth discussion of the effective radius, $R_{\nu}$.

A numerical approach allows for these assumptions to be relaxed and for the acceleration region to be included in the calculations. For this to be accomplished, the wind density profile must be specified according to the velocity profile given by the results of CAK theory. The following section describes this process.

\subsection{Accelerating Stellar Wind} 

\label{sec:General_radio}

The theory set out below follows closely the method used by \cite{Stevens1995}. The formulation of the problem that is presented here begins in a 3D Cartesian geometry. Refinement of the model then reduces it to cylindrically geometry. This is accomplished by first generating a 3D grid and assigning a value of density to each grid point. For the simple case of a spherically symmetric, monotonically increasing velocity wind, the density is given by:
\begin{equation}
	n_{\mathrm{i}} = \frac{\dot{M}}{ 4 \pi \mu_{i} m_{\mathrm{H}} r^{2} v(r) } = \frac{A}{r^{2}}
	\label{eq:density}
\end{equation}
where $\mu_{i}$ is the mean mass per ion in (amu), $m_{H}$ is the proton mass and $n_{\mathrm{i}}$ is the wind ion density. This density profile comes directly from mass conservation and is a general result for a stellar wind. What makes the density profile specific to a particular star is the form which $v(r)$ takes. In the case of a massive star with a CAK wind, $v(r)$ can be represented by: 
\begin{equation}
	v(r) = v_{\infty} \left( 1 - R_{\ast}/r \right)^{\beta},
	\label{eq:vel_law}
\end{equation}
where $\beta$ determines the steepness of the velocity profile. A large $\beta$ value leads to a more gradual acceleration and vice versa. Hence, larger values of $\beta$ allow the acceleration region of the star to protrude further into the wind. 

We have assumed this velocity law, however there are numerous other velocity laws which could have been employed, see \cite{Muller2008} for an in depth comparison of alternatives to equation (\ref{eq:vel_law}). In addition, it is now thought that massive have a small sub-surface convection zone \citep{Cantiello2009} and this region may well be responsible for generating the perturbations that give rise to clumping in the wind (see Section \ref{sec:Clumping_theory}).

Equation (\ref{eq:vel_law}) assumes spherical symmetry, deviations from this would affect the resulting radio flux, however the effect is not usually large. For simplicity we ignore non-spherical symmetry in this work (see \citealt{Schmid-Burgk1982} for a discussion of non-spherical symmetry for terminal velocity wind models and simple geometries).

Defining the Cartesian coordinates $x$, $y$ and $z$, where $x$ is the direction of the line of sight of the observer (where the observer is situated at $x~=~+\infty$). The total intensity $I_{\nu}(y,z)$ in terms of the Planck function $B_{\nu} \left[ T(x,y,z) \right]$ and the optical depth $\tau (x,y,z)$ is:
\begin{equation}
        I_{\nu}(y,z) = \int_{-\infty}^{+\infty} B_{\nu} \left[ T(x,y,z) \right] \exp(-\tau (x,y,z)) \kappa_{\mathrm{ff}} (x,y,z) \mathrm{d} x,
	\label{eq:intensity}
\end{equation}
in which $\kappa_{\mathrm{\mathrm{ff}}} (x,y,z)$ is the free-free absorption coefficient given by $\kappa_{\mathrm{\mathrm{ff}}}~=~\kappa_{\mathrm{\mathrm{e}}} \rho$ where $\kappa_{\mathrm{e}}$ is the electron scattering opacity.

The infinitesimal optical depth of the wind material across the distance $\mathrm{d}x$ can be defined as $\mathrm{d} \tau~=~-~\kappa_{\mathrm{ff}}(x,y,z)~\mathrm{d}x $, where the negative symbol indicates that $\tau$ decreases from the observer to the point of emission \citep{Stevens1995}. Substitution of this expression into equation (\ref{eq:intensity}) allows for it to be recast in terms of the maximum optical depth, $\tau_{\mathrm{max}}$, along the observers line of sight, which results in the line of sight intensity
\begin{equation}
        \label{eq:I_iso}
        I_{\nu}(y,z) = B_{\nu} \left( T \right) \int_{0}^{\tau_{\mathrm{max}}(y,z)} \exp(-\tau (x,y,z)) \mathrm{d} \tau,
\end{equation}
where the isothermal assumption has been applied to allow the Planck function to be removed from the integrand. Integration of equation (\ref{eq:I_iso}) leads to
\begin{equation}
        \label{eq:I_tau_max}
        I_{\nu}(y,z) = B_{\nu} \left( T \right) \left[1-\exp(-\tau_{\mathrm{max}}(y,z)) \right],  
\end{equation}
therefore $I_{\nu}(y,z)$ is only a spacial function of $\tau_{\mathrm{max}}(y,z)$, along each column in $x$. At radio and sub-mm frequencies $h v \ll k_{\mathrm{B}} T$, leading to $B_{\nu}(T) \sim 2 k_{\mathrm{B}} T_{\mathrm{eff}} \nu^{2} / c^{2}$. where $h$ is Planck's constant. At this point the wind temperature, $T$, has been replaced by the stellar effective temperature, $T_{\mathrm{eff}}$, an assumption we make for the remainder of this work. 

Under the same condition which allows the Planck function to be simplified, $\kappa_{\mathrm{\mathrm{ff}}} (x,y,z)$ can be re-expressed as a function of frequency and temperature such that
\begin{equation}
        \label{eq:kappa_nu_t}
        \kappa_{\nu}(T_{\mathrm{eff}}) = 0.0178 \frac{Z^{2} \textit{\textg}_{\mathrm{ff}} }{T_{\mathrm{eff}}^{3/2} \nu^{2}} n_{\mathrm{e}} n_{\mathrm{i}} = K_{\nu}(T_{\mathrm{eff}}) n_{\mathrm{e}} n_{\mathrm{i}}.
\end{equation}
This expression retains its spacial dependence due to $n_{\mathrm{e}}$ and $n_{\mathrm{i}}$, which are the local electron and ion number densities respectively. \textit{\textg}$_{\mathrm{ff}}$ is the free-free Gaunt factor (see Section \ref{sec:gaunt}) given by:
\begin{equation}
	\mathrm{\textit{\textg}}_{\mathrm{ff}} = 9.77 + 1.27 \log_{10} \left( \frac{T_{\mathrm{eff}}^{3/2}}{\nu Z} \right)
	\label{eq:gff}
\end{equation}
\citep{Stevens1995}. 

The number densities $n_{\mathrm{e}}$ and $n_{\mathrm{i}}$ are related through the ratio $\gamma~=~n_{\mathrm{e}}/n_{\mathrm{i}}~\sim~1$, allowing the electron number density to be removed from the expression and replaced by $n_{\mathrm{e}}~=~\gamma n_{\mathrm{i}}$. The value of $\gamma$ is dependent upon the wind metallicity (which is assumed to be solar). For solar abundances $\gamma$ will be approximately independent of radius as long as H is fully 
ionised. Following from equation (\ref{eq:kappa_nu_t}), $\mathrm{d} \tau = K(\nu,T) \gamma n_{\mathrm{i}}^{2} \mathrm{d} x$, allowing $\tau_{\mathrm{max}}$ in equation (\ref{eq:I_tau_max}) to be written as:
\begin{equation}
        \label{eq:tau_continus}
        \tau_{\mathrm{max}}(y,z) = \int_{-\infty}^{+\infty} \gamma K_{\nu}(T_{\mathrm{eff}}) n_{\mathrm{i}}^{2}(x,y,z) \mathrm{d} x.
\end{equation}
Bringing together equations (\ref{eq:flux_full}) and (\ref{eq:I_tau_max}) gives the total flux, 
\begin{equation}
        \label{eq:flux_contiuns}
        S_{\nu} = \frac{B_{\nu} \left( T_{\mathrm{eff}} \right)}{D^{2}} \int_{0}^{\infty} \left[1-\exp(-\tau_{\mathrm{max}}(y,z)) \right] \mathrm{d}y \mathrm{d}z,
\end{equation}
from the wind.

We have assumed $T = T_{\mathrm{eff}}$, the flux $S_\nu \propto B_\nu K_\nu^{2/3}$. Here $B_\nu \propto \nu^2 T_{\mathrm{eff}}$ and $K_\nu~\propto~\textit{\textg}_{\mathrm{ff}}\nu^{-2} T_{\mathrm{eff}}^{-3/2}$, so that $S_\nu \propto \nu^{2/3} \textit{\textg}_{\mathrm{ff}}^{2/3}$, which is the same result as in WB75. Consequently, $S_\nu$ is largely independent of the assumed value of $T$ (assuming that the wind remains largely ionised), or rather only has a small $T$ dependence through the Gaunt factor \citep{Schmid-Burgk1982}. Therefore, models with an assumed temperature gradient will yield rather similar results to those presented here.

Together, equations (\ref{eq:tau_continus} and \ref{eq:flux_contiuns}) allow for the calculation of free-free thermal radio emission from a density distribution, $\rho(r)$. However, these equations are still continuous and in order for them to be applied to a discrete density grid, they need to be discretized. This is a trivial step and involves simply replacing the integrals over the three Cartesian coordinates, $x$, $y$ and $z$, with summations. 

Moving from Cartesian to cylindrical geometry reduces the computational demand of the calculation. This is done by dropping the dependence upon $y$, and the summation over that coordinate and then multiplying by $2 \pi z$, where $0 < z < z_{\mathrm{max}}$. Both the maximum optical depth and total flux are then written in cylindrically symmetric, numerical form as:
\begin{equation}
        \label{eq:num_tau_cly}
        \tau_{\mathrm{max}}(z) = \gamma K_{\nu}(T_{\mathrm{eff}}) \sum_{x} n_{\mathrm{i}}^{2}(x,z)
\end{equation}
and
\begin{equation}
        \label{eq:num_flux_cly}
        S_{\nu} = \frac{2 \pi B_{\nu} \left( T_{\mathrm{eff}} \right)}{D^{2}} \sum_{z} \left[1-\exp(-\tau_{\mathrm{max}}(z)) \right] z.
\end{equation}
These two equations form the final expressions that were used to generate the results presented in Section \ref{sec:Res}.

\subsection{Effective Radius and Acceleration}
\label{sec:effect_radius}

WB75 defined the characteristic radius by taking the point at which the free-free optical depth $\tau_{ff}=0.244$. The physical meaning of this characteristic radius (discussed in Section \ref{sec:radio_emission_symmetry}) is rather vague (and has often been over-interpreted) and other characteristic radii can be defined. For example \cite{Panagia1975} define their characteristic radius as that radius from within which half the free-free flux originates.

The WB75 characteristic radius is defined as the radius where, for all radii greater than this, the wind material is optical thin and contributes to the total emission. The flux of optically thin free-free emission from radii in this range ($R_\nu~<~r~<~\infty$) is: 
\begin{equation}
F_\nu =\frac{1}{D^2}\int_{R_\nu}^\infty  4\pi r^2 j_{ff}(r) dr
\end{equation}
with 
\begin{equation} 
j_{ff}=1.4\times 10^{-27}T_{\mathrm{eff}}^{1/2} n_e n_i g_{ff}.
\end{equation}

A more useful definition, suggested by \cite{vanloo2004}, is to define a characteristic radius as that radius where $\tau_{ff}~=~1$ (integrating from infinity down to a radius $R_\nu$). For a terminal velocity, spherically symmetric wind this is easy to calculate. Therefore, the free-free optical depth 
from $\infty$ to an effective radius $R_{\nu}$ is
\begin{equation}
\tau_{ff} = \gamma K_{\nu}(T_{\mathrm{eff}}) A^2 \int_{R_\nu}^\infty \frac{1}{r^4} dr = \frac{\gamma K_{\nu}(T_{\mathrm{eff}}) A^2}{3R_\nu^3}. 
\end{equation}
Rearranging for $R_{\nu}$:
\begin{equation}
R_\nu= \left[\frac{\gamma K_{\nu}(T_{\mathrm{eff}}) A^2}{3 \tau_{ff}}\right]^{1/3}.
\label{eqn:non_accel}
\end{equation} 
The radius ratio between models with $\tau_{ff}~=~1.0$ and $\tau_{ff}~=~0.244$ (i.e. the WB75 characteristic radius) is then $R_\nu(\tau_{ff}=1)/R_\nu(\tau_{ff}=0.244)~=~0.62$, Bringing $R_{\nu}$ closer to $R_{\ast}$ for $\tau_{ff}~=~1$. Expressing this in convenient units we have that
\begin{equation}
R_\nu(\tau_{ff}=1)=1.75\times 10^{28} Z^{2/3} g_{ff}^{2/3}
T^{-1/2}\left(\frac{\dot M}{\mu v_\infty \nu}\right)^{2/3} 
\end{equation}
with $\dot M$ in $M_\odot$ yr$^{-1}$ and $v_\infty$ in km~s$^{-1}$.

Using model S1 as a representable set of stellar parameters, at an observing frequency of, 600\,GHz, we have $R_\nu(\tau_{ff}~=~1)~=~1.7 \ R_\ast$ and at 900\,GHz, $R_\nu(\tau_{ff}~=~1)~=~1.3 \ R_\ast$. For a $\beta=0.8$ wind acceleration model, we are already clearly deep in the wind acceleration zone. At this frequency, for models with lower mass-loss rates, a terminal velocity wind may lead to a characteristic radius smaller than the stellar radius. Meaning that wind acceleration must be accounted for within the ALMA bands. However, this is only true for a constant velocity model, $R_\nu(\tau_{ff}~=~1)$ is a general quantity and can easily be calculate in the case of an accelerating wind. There are analytic solutions for the free-free optical depth for some $\beta$ wind velocity models (equation (\ref{eq:vel_law})). We can find solutions for $\beta~=~1/2$, $\beta~=~1$, $\beta~=~3/2$, $\beta~=~2$ and so on. In this case we have the nucleon density 
\begin{equation}
n_{\mathrm{i}} = \frac{A}{r^2 (1-R_\ast/r)^\beta}
\end{equation}
and the free-free optical depth from infinity to a radius $R_\nu$ given by
\begin{equation}
\tau_{ff}= \gamma K_{\nu}(T_{\mathrm{eff}}) \int_{R_\nu}^\infty n_{\mathrm{i}} dr = (\gamma K_{\nu}(T_{\mathrm{eff}}) A^2) I(\beta)
\label{eqn:tauff}
\end{equation}
where $I(\beta)$ is an integral whose solution specifically depends on the velocity law. When $\beta=0$ we have $I(0)~=~1/(3R_\nu^3)$, giving the earlier expression of equation (\ref{eqn:non_accel}).

For $\beta~=~2$ we have the following integral
\begin{equation}
I(\beta=2)= \int_{R_\nu}^\infty \frac{dr}{r^2(1-R_\ast/r)^{2}} = \frac{1}{3}\left[\frac{1}{(R_\nu-R_\ast)}\right]^3
\label{eqn:b2}
\end{equation}
which reduces to the constant velocity result when $R_\nu~\gg~R_\ast$ (which is obviously the case). Substituting equation (\ref{eqn:b2}) into equation (\ref{eqn:tauff});
\begin{equation}
\tau_{ff} = (\gamma K_{\nu}(T_{\mathrm{eff}}) A^2) \frac{1}{3}\left[\frac{1}{(R_\nu-R_\ast)}\right]^3
\end{equation}
and rearranging gives 
\begin{equation}
R_\nu = \left[ \frac{\gamma K_{\nu}(T_{\mathrm{eff}}) A^2}{3 \tau_{ff}} \right]^{1/3} + R_\ast.
\label{eqn:R_asum}
\end{equation}

\begin{figure}
\centering
\includegraphics[width=0.5\textwidth]{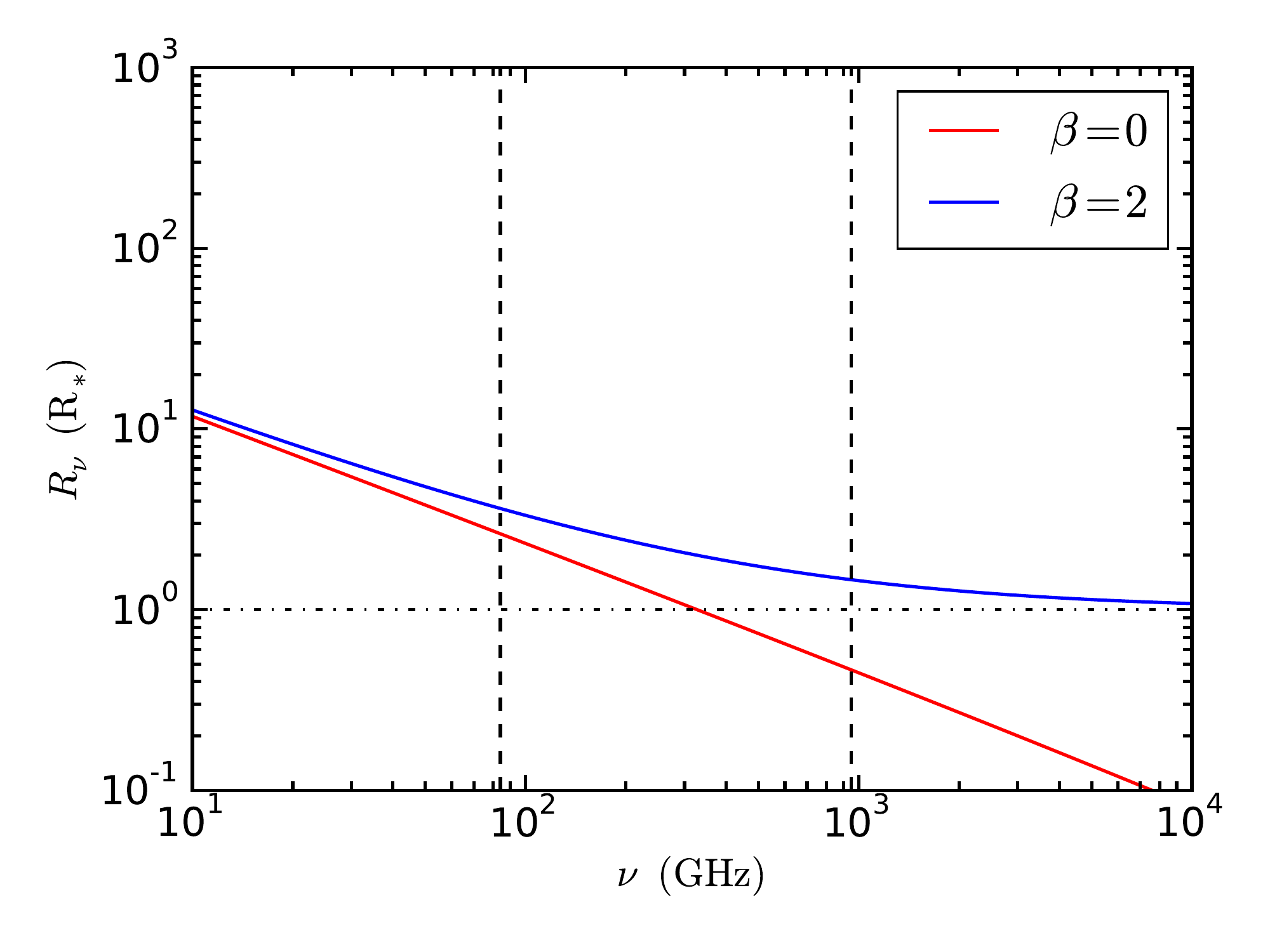}
\caption{
Effective stellar radius as a function frequency for velocity laws with $\beta~=~0$ and $\beta~=~2$. $\beta~=~0$ corresponds to the WB75 model where $S_{\nu}~\propto~\nu^{0.6}$. As $\beta$ increases there is deviation from this linear model as in the case of $\beta~=~2$. All non zero values of $\beta$ are asymptotic to $R_{\nu}~=~1$, only for a velocity law where $\beta~=~0$ dose $R_{\nu}$ value less than $R_{\ast}$. The vertical dashed lines indicate the ALMA frequency range for bands 3 - 9.
\label{fig:R_nu}
}
\end{figure}

Equation (\ref{eqn:R_asum}) is identical to equation (\ref{eqn:non_accel}) with the addition of an extra term on the right hand side, $R_{\ast}$. This accounts for the characteristic radius being asymptotic to $R_{\ast}$ at higher frequencies ($\nu~>~100$ GHz). Fig. \ref{fig:R_nu} shows $R_{\nu}$ for both $\beta~=~0$ and $\beta~=~2$ with $\tau_{ff}~=~1$. 

For model S1, at an observing frequency of $\nu~\approx~900$ GHz, the WB75 model results in $R_{\nu}~\approx~0.5 R_{\ast}$ which is obviously non-physical, however the accelerating wind model results in $R_{\nu}~\approx~1.7 R_{\ast}$, which has physical meaning. This difference shows the importance in accounting for the wind acceleration region at high radio and sub-mm wavelengths.

\subsection{Gaunt Factor}
\label{sec:gaunt}

A recent study by \cite{vanHoof2014} has improved the understanding of \textit{\textg}$_{\mathrm{ff}}$, by performing calculations at unprecedented accuracy and across a larger parameter space than has thus far been attempted. \cite{vanHoof2014} communicate an extensive data set, allowing testing of the equation (\ref{eq:gff}). \cite{vanHoof2014} provide values of the Gaunt factor as a function of $u = h \nu / k_{\mathrm{B}} T_{\mathrm{e}}$ and $\gamma = Z^{2} \mathrm{Ry} / k_{B} T_{\mathrm{e}}$. Here $T_{\mathrm{e}}$ is the electron temperature and Ry is the infinite-mass Rydberg unit of energy (13.606~eV). Under the isothermal assumption $T_{\mathrm{e}} = T_{\mathrm{eff}}$, for $T_{\mathrm{eff}} = 40 \times 10^{3}$~K, $u$ falls within the range $10^{-5} < u < 10^{-2}$.

\begin{figure}
\centering
\includegraphics[width=0.5\textwidth]{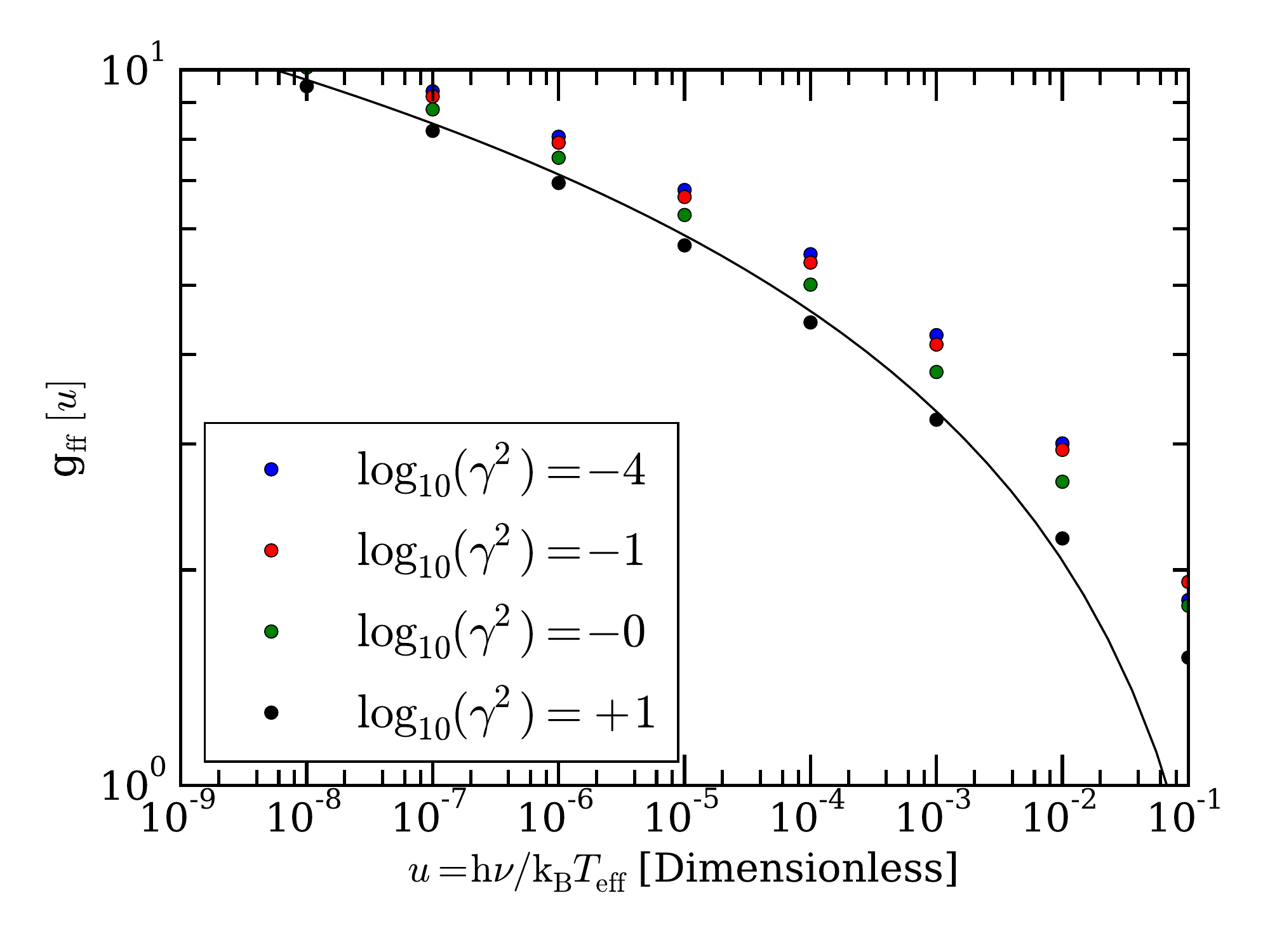}
\caption{
The free-free Gaunt factor, \textit{\textg}$_{\mathrm{ff}}$, as a function of the parameter $u = h \nu / k_{\mathrm{B}} T_{\mathrm{e}}$ for several values of $\gamma$. The solid line represents equation (\ref{eq:gff}) and the data are from \protect\cite{vanHoof2014}. 
\label{fig:gff}
}
\end{figure}

Fig. \ref{fig:gff} plots a selection of results from \cite{vanHoof2014} along with equation (\ref{eq:gff}). Due to the good agreement between the data and equation (\ref{eq:gff}) (with only a slight deviation at higher values of $u$) and the relatively weak dependence of $S_{\nu}$ on \textit{\textg} $_{\mathrm{ff}}$, equation (\ref{eq:gff}) was deemed to be sufficiently accurate for the temperature and frequency ranges that were investigated.

\subsection{Wind Clumping and the Clumping Factor}
\label{sec:Clumping_theory}

Due to the instabilities inherent in line-driven stellar winds \citep{OCR} we expect the winds of massive stars to be clumped and it is highly likely that the degree of clumping will vary strongly with radius. Such a radially varying clumping will have an impact on the spectral shape of the radio/sub-mm emission from massive stars. Clumping has been discussed in the literature already, for example, WB75, \citet{Abbott1981} and \citet{vanloo2004,vanloo2006} amongst others.

If the clumping is constant throughout the wind, for a specific mass-loss rate, the flux will be raised uniformly across all wavelengths. This means that there will be an over estimate of the mass-loss rate. If the clumping is not constant then the effect will be different at different wavelengths, allowing the possibility of a more detailed investigation of how clumping varies radially.

Clumping has also been included in spectral modelling of the optical/IR part of the spectrum, for example \citet{Hillier1999} and \citet{Oskinova}.

\citet{Oskinova} discussed the difference between microclumping and macroclumping. The basic assumption of microclumping is that the wind clumps are small compared to the mean free path of photons (i.e. optically thin). In spectral lines this may not be true (where the line centre can have a large optical depth). Where clumps are optically thick (or optically thick at some frequencies) this is generally referred to as \lq\lq macroclumping\rq\rq\ and here concepts of porosity in the wind come into play (see \citealp{Owocki} for a discussion of the possible impact of macroclumping and porosity on the X-ray line profiles of massive stars).

\citet{Ignace} has discussed the consequences of different forms of macroclumping on the expected radio/sub-mm spectra of the winds of massive stars. Here we focus on microclumping and specifically radially varying microclumping.

For the continuum free-free processes it is likely to be the case that we are dealing with microclumping for much of the wind. Remember though that the free-free opacity is a strong function of frequency and density ($\propto (n_i n_e)/\nu^2$) and so a clump that is optically thin at high frequencies will become optically thick at low frequencies.
 
For the clumped wind calculations presented here, we assume that the wind consists of small, optically thin clumps. These clumps fill a volume $f_V$ and an
inter clump medium that is essentially a void. This is clearly a major assumption and we would expect a range of clump densities at all radii. Following on from \citet{Runacres2002}, the clumping factor $f_{cl}$ is defined as 
\begin{equation}
f_{cl}=\frac{<\rho^2>}{<\rho>^2}=\frac{1}{f_V}\,.  
\end{equation} 
Where $\rho$ is the wind density and the symbol $<>$ is the time average of the quantity. In this case $f_{cl}>1$ and $f_V<1$. Optical analyses have suggested quite large values of $f_{cl}~\sim~10-50$ in the inner wind, where the optical lines are formed \citep{Crowther2002,Bouret2005}. However, clumping factors of 50 would have a very large effect on the results and it is unlikely that such large clumping factors are correct, see \cite{Oskinova}, as such we have not used factors of this magnitude in this work. The maximum clumping factor used here is $f_{cl}=2$.

In general, for case were $f_{cl}$ is constant throughout the wind, then the flux $S_\nu$ from a clumped wind scales as $S_\nu\propto (\dot M^2 f_{cl})^{2/3}$, so that the presence of clumping reduces the mass-loss required to give the same level of emission.

There are several different versions for the assumed form of clumping. \citet{Hillier1999,Bouret2005} used this form for the {\em volume filling} clumping factor in calculations using the CMFGEN code:
\begin{equation}
f_V(r) = f_{V,\infty} + (1-f_{V,\infty})e^{-v(r)/v_{cl}} 
\end{equation}
where $v(r)$ is the wind velocity as a function of radius $r$ (For all clumped wind models present here, $v(r)$ is assumed to have the form of equation (\ref{eq:vel_law}) with $\beta~=~0.8$) and $v_{cl}$ is the velocity at which clumping starts. The parameter $f_{V,\infty}$ is effectively the volume filling clumping factor at large radii.

Typical values discussed in terms of fitting optical line profiles are $f_{V,\infty}=0.1$ and $v_{cl}=30$~km~s$^{-1}$ \citet{Bouret2005}, i.e. clumping starts very close to the star and there is a very significant degree of clumping. Also, in this model the wind stays clumped out to large radii, once clumped the wind does not become unclumped. It is worth noting that optical line profiles do not seem to be that sensitive to clumping and are formed closer to the star, as compared to the bulk of the radio/sub-mm emission.

\begin{figure} 
\centering
\includegraphics[width=0.5\textwidth]{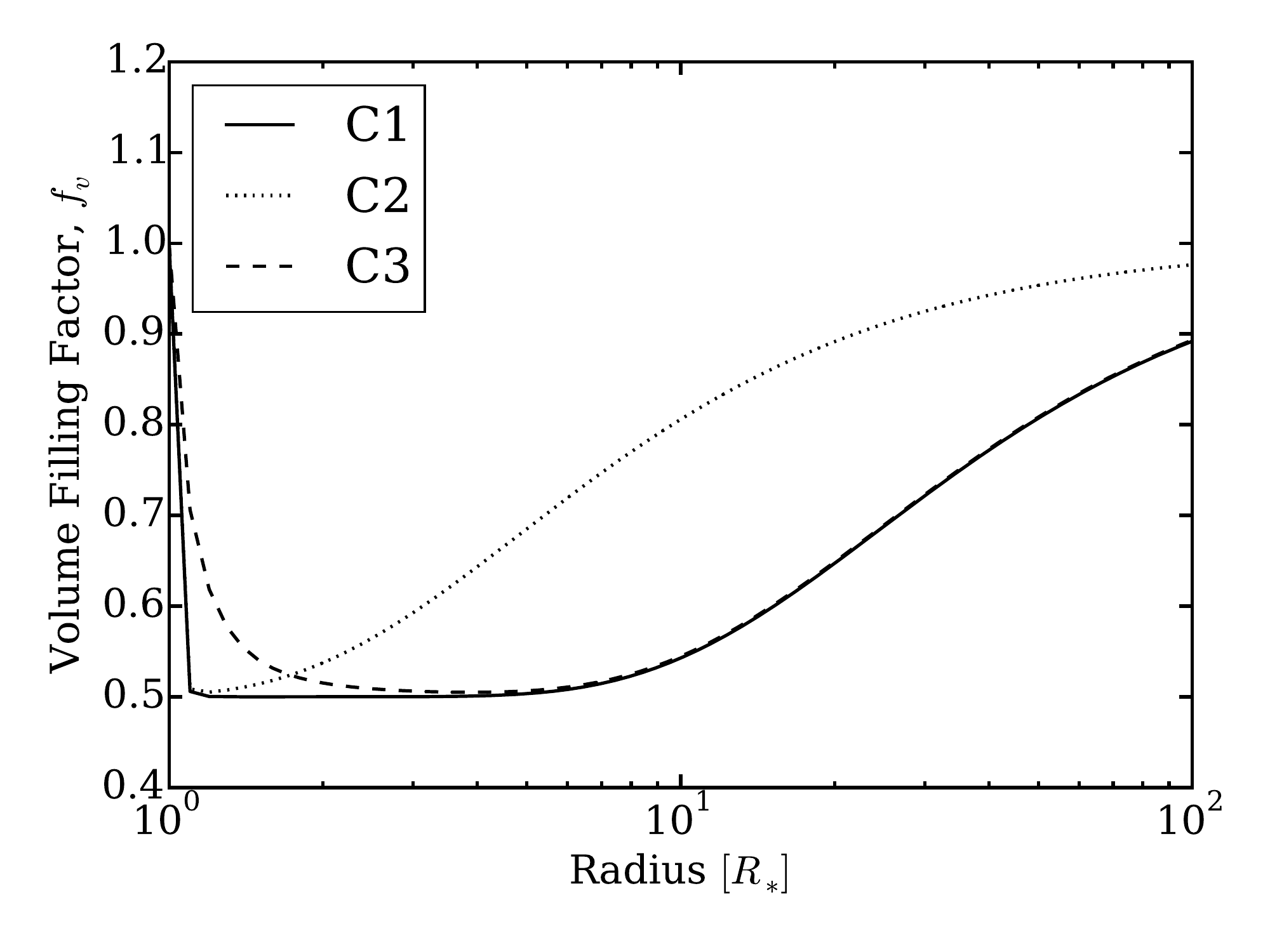} 
\caption{
The radially dependent volume filling clumping parameter for the models set out in Table~\ref{tab:tab_cl}. The uniform clumping models have $f_V=1$ for the smooth model (Model S2) and $f_V=0.5$ for the uniform clumping model (Model C0). neither of these lines are shown for purposes of clarity.
\label{cl1}
}  
\end{figure}

Related to this, \citet{Schnurr2008} have suggested another prescription for the clumping, namely
\begin{equation} 
f_V(r) = f_{V,max} + (1-f_{V,max})\left[e^{-v(r)/v_1} +
e^{(v(r)-v_\infty)/v_2}\right] 
\end{equation} 
where $v_1$ and $v_2$ are constants and $f_{V,max}$ sets the minimum value of the volume filling clumping factor (i.e. maximum clumping). In contrast to the previous case, $f_V(r)\rightarrow 1$ as $r\rightarrow \infty$. We adopt this clumping law here and note that it resembles (in very broad outline only) the results from the theoretical calculations of \citet{Runacres2002}. We have not attempted to adjust the clumping parameters ($v_1$, $v_2$, $f_{V,max}$) to reproduce their results. We note that the clumping in the \citet{Runacres2002} results is rather more pronounced, with a peak clumping parameter $f_{cl}> 10$, and with the peak clumping occurring at around $10-20R_\ast$ (at rather larger radii than assumed here). The radially varying volume filling clumping factors are shown in Fig.~\ref{cl1}, excluding the constant clumping models for sake of clarity. Table \ref{tab:tab_cl} shows the clumped wind models C0 - C3 used in this study, which all have an underlying smooth wind model given by model S2. 

With this parameterisation of clumping we can recalculate the expected radio/sub-mm emission, using the prescription described earlier, using the clumped values of the density in each cell, with $\rho_{cl}=\rho_{sm}/f_V$, with $\rho_{cl}$ the clumped density at a given radius, $\rho_{sm}$ the smooth density at a given radius and $f_V$ as above. In addition, in each cell the clumps comprise only a fraction $f_V$ of the path, the rest being a void. 

\begin{table}
\caption{The parameters used for the clumping model}
\centering
\begin{tabular}{lllll}\hline
Model & $f_{max}$ & $v_1$ & $v_2$ & Comment\\\hline
S2 & 1.0 & 0   &  0  & No clumping\\
C0 & 0.5 & 0   &  0  & Uniform clumping\\
C1 & 0.5 & 100 & 100 & Radially dependent clumping\\
C2 & 0.5 & 100 & 500 & Less clumping at large $r$\\
C3 & 0.5 & 500 & 100 & Less clumping at small $r$\\\hline
\end{tabular}
\label{tab:tab_cl}
\end{table}

\subsection{Numerical Setup}

A grid was initialised for $\{x_{i},z_{j}\}$ with $x_{i_{\mathrm{max}}} = 2 \times 10^{5}$, approximately covering the range $-1800 \ R_{\ast} < x < 1800 \ R_{\ast}$, and $z_{j_{\mathrm{max}}} = 1 \times 10^{5}$, approximately covering the range $0 < z < 1800 \ R_{\ast}$. This gives and inter-grid spacing $\Delta x~=~\Delta z = 0.018 \ R_{\ast}$. This grid covers a 2D spatial extent, which encompasses the outer regions of the stellar wind while still having the necessary resolution to resolve the flux from close to the stellar surface. The star is located at the origin of the coordinate system, at the centre of the lower edge of the domain. Using equations (\ref{eq:density}) and (\ref{eq:vel_law}), the grid is populated with ion number densities.

The calculations are performed according to the theory defined in Section \ref{sec:General_radio}; the maximum optical depth, $\tau_{\mathrm{max}}(z)$, for each column in $x$ (along the line of sight) is calculated using equation (\ref{eq:num_tau_cly}). Then summing all values of $\tau_{\mathrm{max}}$ with equation (\ref{eq:num_flux_cly}) gives the total flux $S_{\nu}$. The region where $r < R_{\ast}$ was set to an arbitrarily large value, such that it appears as an optically thick medium. Finally, $S_{\nu}$ is calculated according to the WB75 method (i.e assuming a terminal velocity wind) using equation (\ref{eq:total_flux}), which we shall refer to as $S_{\nu,WB75}$. The next section will present, compare and discuss the results of these calculations. 

\section{Results and Discussion}
\label{sec:Res}

\begin{figure}
\centering
\includegraphics[width=0.5\textwidth]{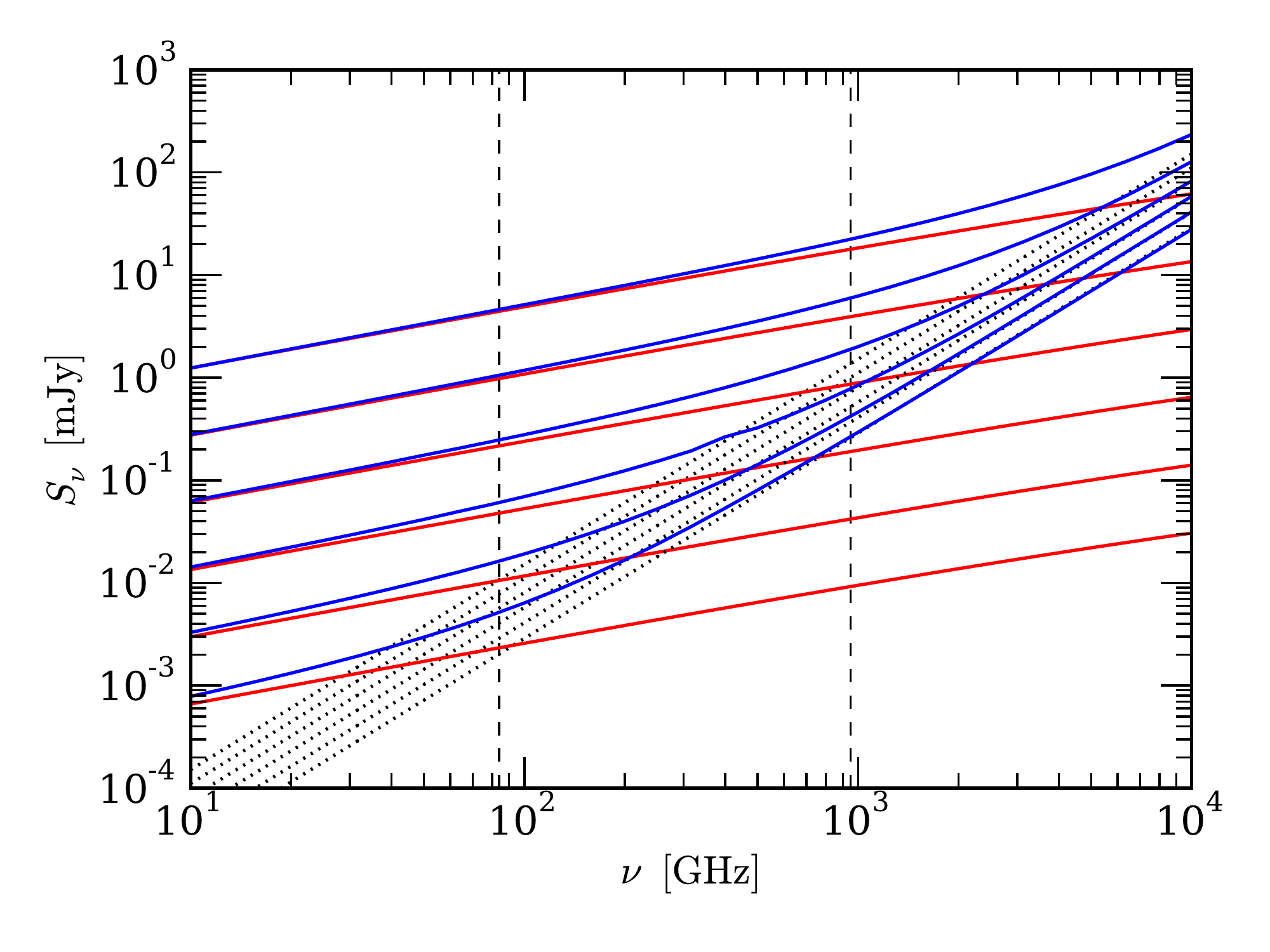}
\caption{
The predicted radio flux, $S_{\nu}$, from our ensemble of stellar models, assuming a distance of $0.5$~kpc and wind velocity parameter $\beta~=~0.8$. The results are shown as a function of $\nu$ for stellar models S0 - S5. The thermal radio emission results from the terminal velocity WB75 model are shown as the red lines. The numerical results from the new accelerating wind models are the blue lines. The black dotted lines are the RJ curves of the emission from the stellar surface. Each separate set of lines corresponds to a different value of $\dot{M}$, increasing from bottom to top as the mass-loss rate increases. The vertical dashed lines indicate the ALMA frequency range for bands 3 - 9.
\label{fig:spherical_wind_M_dot}
}
\end{figure}

\begin{figure*}
\centering
\includegraphics[width=0.84\textwidth]{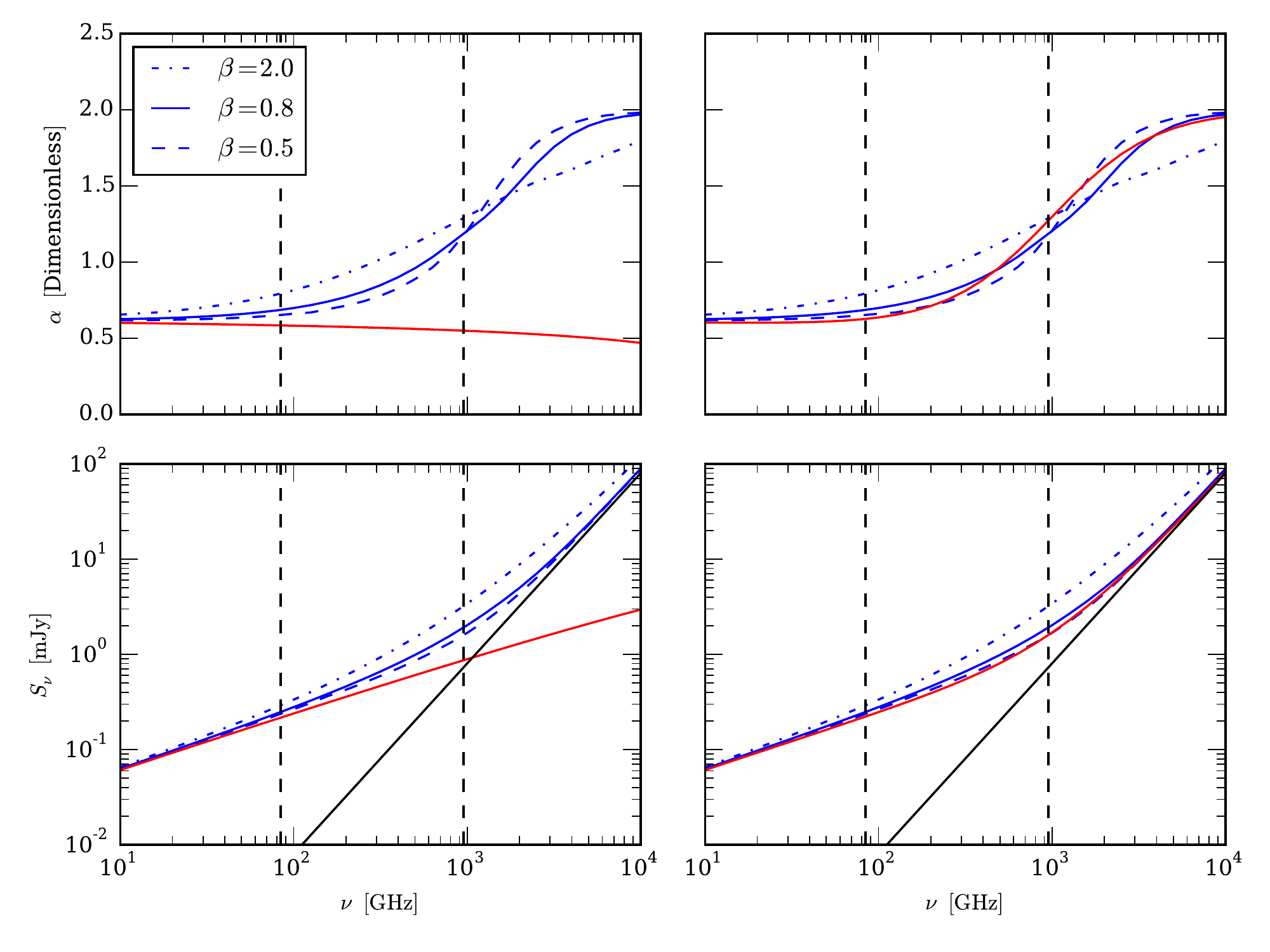}
\caption{
The predicted radio flux, $S_{\nu}$ (bottom row of diagrams), and spectral index, $\alpha$ (top row of diagrams), for wind model S2. The red lines indicate the result from the WB75 model and the blue lines indicate the results for the numerical model. There are three separate values for the velocity law parameter: $\beta = 0.5$, $\beta = 0.8$ and $\beta = 2.0$, indicated by the dashed line, the solid line and the dot-dashed line respectively. The straight black lines indicate the RJ curve of the stellar surface. The right hand column is distinct from the left due to the RJ curve having been added to the WB75 result. The vertical dashed lines approximately indicate the ALMA frequency range. The vertical dashed lines indicate the ALMA frequency range for bands 3 - 9. 
\label{fig:spherical_wind_beta}
}
\end{figure*}

Fig. \ref{fig:spherical_wind_M_dot} presents $S_{\nu}$ for models S0 - S5 calculated using equations (\ref{eq:num_tau_cly}) and (\ref{eq:num_flux_cly}). For each $\dot{M}$ there is clear deviation from the $S_{\nu}$ predicted by WB75. The numerical calculations initially agree with the analytic theory for $\nu < 10^{2}$~GHz. Beyond this, the gradient increases and eventually merges with the Rayleigh-Jeans (RJ) part of the stellar blackbody spectrum for $\nu > 10^{3}$~GHz. This can be seen clearly in the upper right panel of Fig. \ref{fig:spherical_wind_beta} which shows $\alpha$ as a function of $\nu$. Initially $\alpha = 0.6$, corresponding to the WB75 prediction. As the observing frequency increases, $\alpha$ increases ever more rapidly before levelling off with $\alpha \rightarrow 2$ for $\nu > 3 \times 10^{3}$~GHz. The transition between these two values encompasses both the effect of the acceleration region and the point at which the RJ tail begins to contribute significantly to the total spectral flux.

The observing frequency at which the acceleration region begins to contribute is sensitively dependent upon $\dot{M}$. A sufficiently low $\dot{M}$ leads to the acceleration region contributing to $S_{\nu}$ at a flux level which is unobservable by ALMA (if the star is at approximately $0.5$~kpc). This is the case for the lower end of the $\dot{M}$ range, for example the bottom curve of Fig. \ref{fig:spherical_wind_M_dot}. For these curves, the RJ contribution to the spectral flux is completely dominant in the ALMA frequency range. 

In contrast, if an $\dot{M}$ is too high, the acceleration region will contribute to $S_{\nu}$ at a frequency which is higher than the spectral window of ALMA. This does not occur for the $\dot{M}$ range used during this study. The model with largest value of $\dot{M}$, model S0, which corresponds to the top curve in Fig. \ref{fig:spherical_wind_M_dot}, shows a clear divergence from the WB75 prediction and is sufficiently separate from the RJ tail in frequency space to not be effected by its presence. As such, the deviation predicted by the numerical model for the largest value of $\dot{M}$ in this study is due to the acceleration region and is within ALMA's spectral window. This is non obvious in Fig. \ref{fig:spherical_wind_M_dot} due to the logarithmic scale. 

The point at which the gradient begins to change occurs at an increased observing frequency for each increase in $\dot{M}$. The reason for this is that a higher value of $\dot{M}$ leads to larger $R_{\nu}$, requiring a higher observing frequency to see through the wind to the same location. Conversely, at a fixed $\dot{M}$, increasing the observing frequency decreases $R_{\nu}$ until it reaches $R_{\ast}$. However, there is an upper limit on the observing frequency, given by ALMA's spectral window, $\nu < 10^{3}$~GHz. The balance between $\dot{M}$ and observing frequency is key for determining whether the acceleration region of a star's wind is detectable by ALMA.

The physical reasoning behind the increase in $S_{\nu}$ when the acceleration region is taken into account, is that the wind material must undergo acceleration from rest to the terminal velocity. This acceleration leads to a wind density which is greater than would be present if the wind were initial travelling at terminal velocity. Since $\tau_{\mathrm{max}} \propto n_{\mathrm{i}}^{2} \propto \rho^{2}$ and $S_{\nu} \propto [1-\exp(-\tau_{\mathrm{max}})]$, there is a non-linear response from $S_{\nu}$ to a change in the density profile away from the $r^{-2}$ given by the terminal velocity model. This results in the change to $\alpha$ seen in Fig. \ref{fig:spherical_wind_beta}.

The radio flux, $S_{\nu}$, along with the spectral index, $\alpha$, for model S2 are depicted in the bottom two and top two panels of Fig. \ref{fig:spherical_wind_beta} respectively, The left-hand column shows $S_{\nu}$ and $\alpha$ for both the numerical model and the WB75 model, while the right-hand column shows the same, however the RJ part of the stellar blackbody has been added to the WB75 model. This has been done such that the acceleration region is the sole difference between the two methods, allowing for a more direct comparison.

Both the WB75 model and the numerical model show very similar behaviour across the frequency spectrum investigated, with several important distinctions. The plots on the left show the transition between the terminal velocity wind to the stellar blackbody for the numerical model, where by the spectral index, $\alpha = 0.6 \rightarrow 2.0$. The WB75 model experiences no such transition. In fact there is a gradual decrease in $\alpha$, due to the frequency dependence of \textit{\textg}$_{\mathrm{ff}}$. In the left hand column the WB75 model is forced to transit to $\alpha = 2.0$ due to the RJ curve. Here the behaviour of the two models is qualitatively similar. The difference is made apparent by varying $\beta$. As the value of $\beta$ decreases, the density profile becomes steeper, this results in a greater deviation from the WB75 $\alpha$ as the value of $\beta$ decreases. The net effect is to smooth out the acceleration region, leading to the wind passing through a more gradual acceleration, which extends further from the star. Conversely, a smaller $\beta$ leads to a more rapid acceleration which occurs closer to the stellar surface. In the limit of $\beta \rightarrow 0$ the acceleration is instantaneous and the WB75 model is recovered. The value of $\beta$ is therefore an important consideration when predicting $S_{\nu}$. As such, observations of the $S_{\nu}$ across ALMA's spectral window will provide further constraints on the precise value of $\beta$ for a given set of stellar parameters. Leading in turn to a better understanding of $\dot{M}$.

In their analysis, WB75 describe the breakdown of their model at high observing frequencies (or low $\dot{M}$), where sharp temperature gradients and the acceleration region require the solution of the equation of radiative transfer. By using the acceleration law from CAK theory (equation \ref{eq:vel_law}) together with a numerical approach, these complications are avoided.

\begin{figure*}
\centering
\includegraphics[width=0.84\textwidth,trim={0 7cm 0 0},clip]{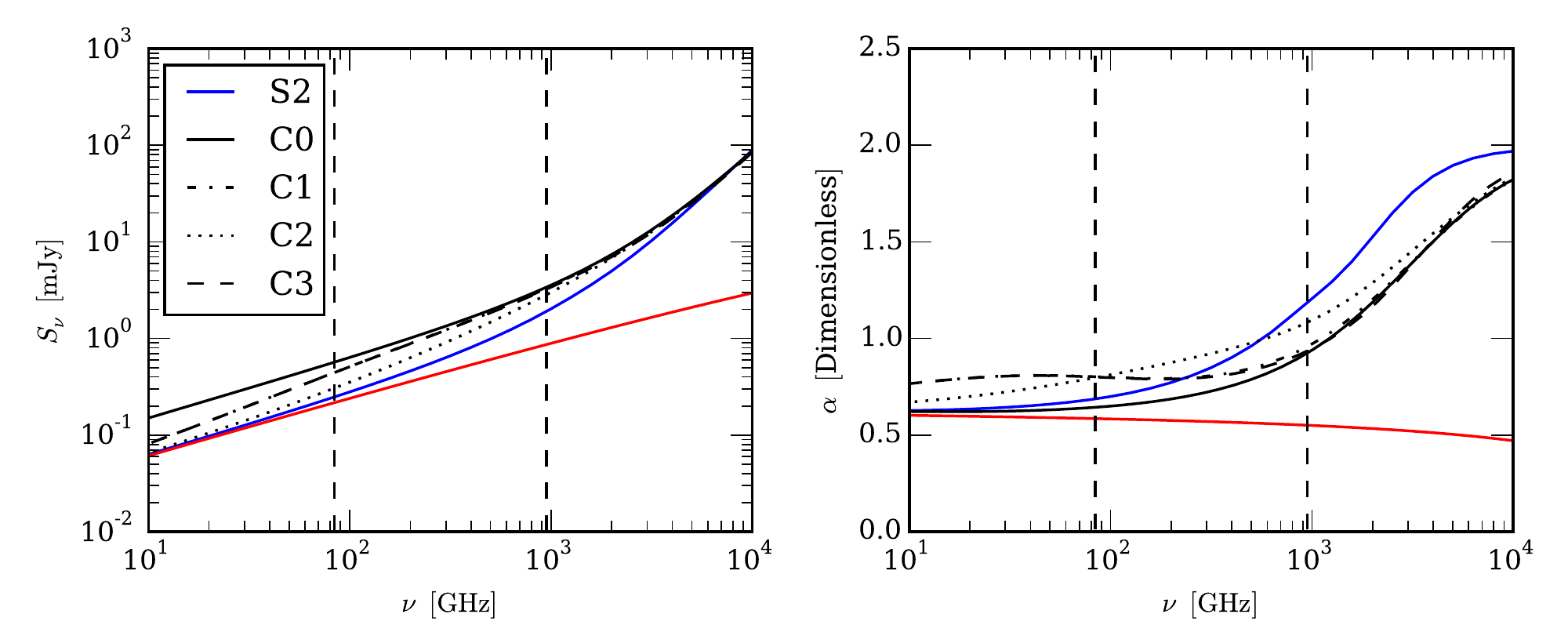}
\caption{
Comparison between clumped wind models C0 - C3 and unclumped wind model S2 (see Tabel \ref{tab:tab_cl}). Both clumped and unclumped models result in an increase in spectral flux for low frequencies ($\nu~\lesssim~500$ GHz). At frequencies higher than this, all models converge to a spectral flux with an index $\alpha = 2$, which is consistent with the RJ blackbody curve. The behaviour of the spectral index $\alpha$ is more nuanced, with the unclumped wind model providing the smoothest transition for $0.6~<~\alpha~<~2$. The clumped wind models take a range of values in the ALMA range (dashed virtical lines). It is notable that clumped wind models C1 and C3 display considerable degeneracy at all frequencies studied, showing only slight divergence at $\nu > 900$ GHz. The vertical dashed lines indicate the ALMA frequency range for bands 3 - 9.
\label{fig:clump_comparison}
}
\end{figure*}

A recent paper by \cite{Manousakis2015} investigates the velocity profile of the X-ray binary Vela X-1. A velocity law with $\beta \sim 0.5$ is found to favour the data. However, the influence of the pulsars radiation field on the dynamics of the donor stars wind are not well understood \cite{Manousakis2015} make reference to work done by \cite{Stevens1993} who investigated different wind velocity laws and the effect on $S_{\nu}$. The treatment of other velocity laws than equation (\ref{eq:vel_law}) is however, beyond the scope of this study. \cite{Thum2013} present millimeter observations of the massive stellar object LkH$\alpha$101. By analysing high-$n$ line transitions they deduce a slow moving wind whose spectral flux corresponds to a non-constant wind velocity. In contrast with the previous analysis, \cite{Blomme2002} attribute observed flux excess from the wind of the early-type star $\eta$ Ori to wind clumping rather than the wind acceleration region. The following section will present the effects of clumping on the results of the spectral flux calculations.

\subsection{Spectral Flux Due to Clumping}

We have calculated the radio/sub-mm spectral flux for several different clumping models, which are summarised in Table~\ref{tab:tab_cl}. All models presented, have wind parameters according to the smooth wind model S2 (see Table~\ref{tab:parameters}). Models C0 - C3 investigate the effects of varying clumping in both the inner and outer wind, as compared to a standard, uniform, clumping model.

For the case of uniform clumping (Model C0), the radio flux is increased by a factor $f_V^{-2/3}$ as expected, and for $f_V(r)=0.5$, the flux is increased by a factor 1.59. The spectral results are shown in Fig. \ref{fig:clump_comparison}, where the flux and the spectral index $\alpha$ are shown.

The clumped models (C0-C3) illustrate how radially varying clumping affects the predicted radio/sub-mm emission from massive stars. As expected, the presence of clumping generally pushes up the expected emission. If the clumping is more pronounced at smaller radii, then the effect on the expected flux is more pronounced at higher frequencies and vice versa (Fig. \ref{fig:clump_comparison}. The changes in the spectral index $\alpha$ are also shown in Fig. \ref{fig:clump_comparison} and these show that substantial changes in $\alpha$ are predicted across the ALMA bands for these clumping models and it will be possible to observe these changes for several nearby O-stars (see below).

\subsection{ALMA Detectability}

ALMA has frequency bands which roughly cover the range $80$~GHz $< \nu <$ $950$~GHz. While previous observations have covered parts of this frequency range, none have so far had comparable sensitivity to ALMA. This sensitivity allows for the detection of sub-mJy flux from an unresolved point source such as the massive stars which are being considered in this work. Therefore, observations which can determine the contribution to $S_{\nu}$ from the acceleration regions of massive stars will be possible providing a further avenue for diagnosis of massive star winds.

\begin{figure}
\centering
\includegraphics[width=0.5\textwidth]{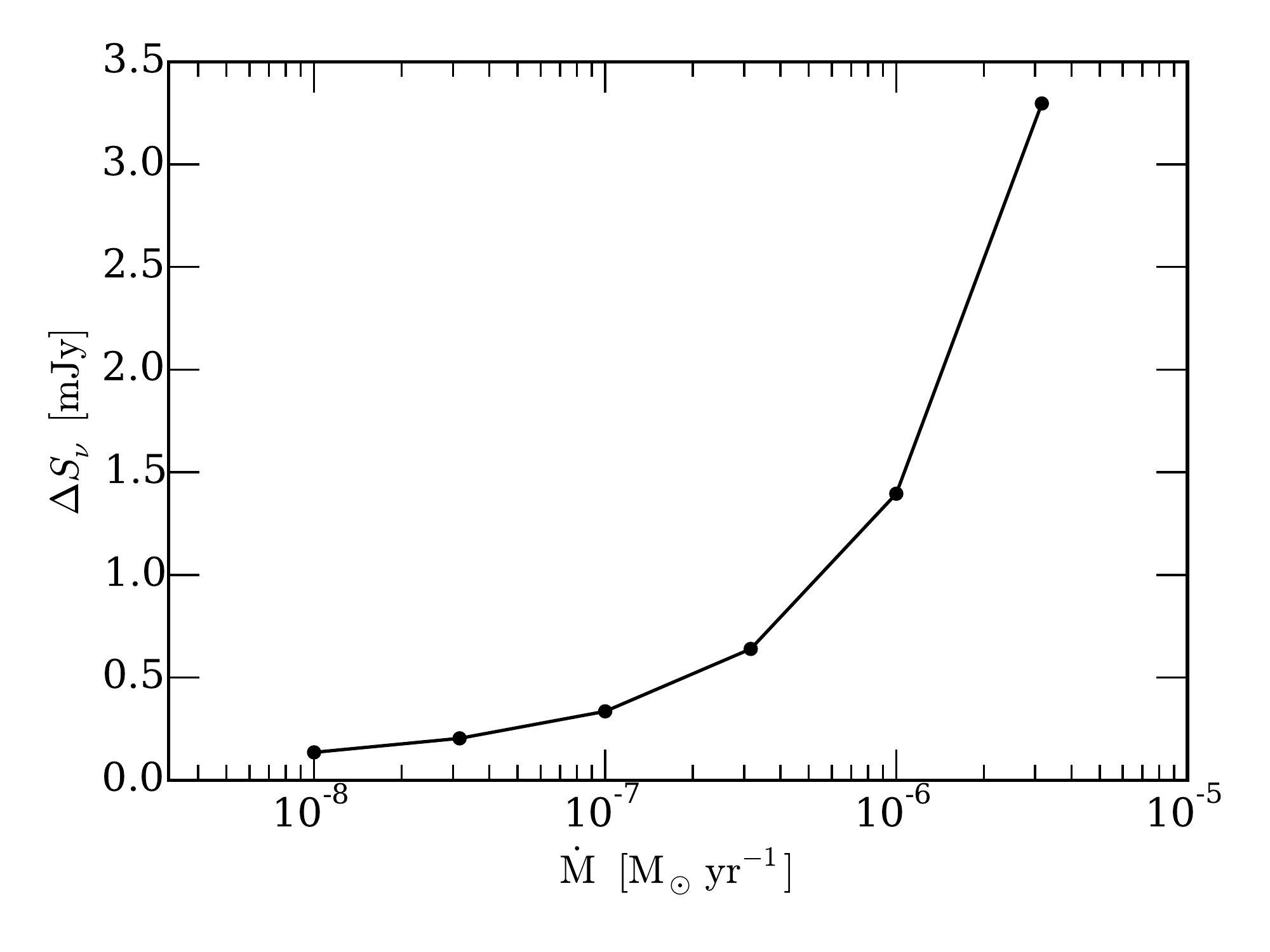}
\caption{
The difference, $\Delta S_{\nu}$, between the accelerating and WB75 models, at a frequency of $\sim 630$~GHz. The effect of the RJ curve has been added to the analytic calculation such that $\Delta S_{\nu}$ is solely due to the contribution to the emission from the acceleration region. 
\label{fig:delta_snu}
}
\end{figure}

To determine the enhancement of $S_{\nu}$ due to the acceleration region we introduce the following quantity: 
\begin{equation}
	\Delta S_{\nu} = S_{\nu,\mathrm{accel}} - S_{\nu,\mathrm{WB75}},
\label{eq:snu_dif}
\end{equation}
where $S_{\nu,\mathrm{accel}}$ is the spectral flux due to the acceleration region and $S_{\nu,\mathrm{WB75}}$ is the spectral flux due to the WB75 model including the RJ flux. Fig. \ref{fig:delta_snu} plots $\Delta S_{\nu}$ at $\nu = 630$~GHz (ALMA Band 9) for all mass-loss rates in the study. As has already been discussed, ALMA's sensitivity is $S_{\nu} \geqslant 0.25$~mJy for an integration time of 1 hour. Mass-loss rates which result in $\Delta S_{\nu} < 0.25$~mJy are therefore not detectable. This point occurs for $\dot{M}~>~10^{-7.5}$~M$_{\sun}$~yr$^{-1}$. Thus a star with $\dot{M}$ larger than this is required for the acceleration region to make an observable contribution to $S_{\nu}$. The largest $\dot{M}$ of this study provides the largest difference in flux: $\Delta S_{\nu} \sim 3.3$~mJy. A spectral flux of this size is detectable by ALMA.

The calculations performed during this work have assumed a distance $D = 0.5$~kpc for each stellar model. Since $S_{\nu} \propto 1/D^{2}$, a more distant object with $\dot{M}~=~10^{-7.5}$~M$_{\sun}$~yr$^{-1}$ will result in a flux lower than the ALMA detectable threshold (at an integration time of 1 hour). Therefore the calculations are most relevant to the study of O type stars with $D < 0.5$~kpc, for example $\zeta$ Pup with $D~\sim~0.33$~kpc or $\zeta$ Oph with $D~\sim~0.15$~kpc \citep{Maiz-Apellaniz2004}.

\section{Conclusions}

Early theoretical work carried out by WB75 on the radio emission from stellar winds of massive stars showed that $S_{\nu} \propto \nu^{0.6}$. This dependence is based on the assumption of a terminal velocity stellar wind. This study has built upon this early work by applying a generalised numerical form of the equations which WB75 began with, to a discrete density grid with a profile that corresponds to the results of CAK theory. As such, this study has been able to investigate the region in the immediate surroundings of a series of stars undergoing mass-loss in the range $10^{-8}$~M$_{\sun}$~yr$^{-1}$ $< \dot{M} <$ $10^{-5.5}$~M$_{\sun}$~yr$^{-1}$. 

Due to a mixture of differing physical regimes within the stellar atmosphere close to the base of the wind, the wind acceleration region is a challenging subject which until recently has received little treatment both theoretically and observationally. This situation is changing due to ALMA and the wind acceleration region has begun to receive attention for example in the context of pre-main sequence stars (see \citealt{Thum2013}).

It has been shown that the spectral index $\alpha$ is strongly non-linear in the ALMA frequency range due to the effect of the wind acceleration region and the gradient strongly depends on the velocity law parameter $\beta$. The excess flux associated with the acceleration region $\Delta S_{\nu}$ 
increases with $\dot{M}$ and should be detectable with ALMA. Therefore, if wind acceleration is not accounted for, miss-identification of the stellar mass-loss rate may occur.

The picture is further complicated by the addition of wind clumping. The effect of clumping on $S_{\nu}$ both at radio and ALMA wavelengths has considerable degeneracy with the smooth wind models. The spectral flux due to a smooth wind model and a specific $\beta$ velocity law may be very similar to a clumped wind model with a different value of $\beta$. Both types of models raise the density with respect to a terminal velocity model, with the precise details varying from model to model. However, we know winds must accelerate, regardless of the details of the clumping posses. Recognising this basic physical property leads directly to acceleration as the baseline for the description of massive star winds at sub-mm frequencies.

The degeneracy between models results in a fundamental ambiguity between the velocity law, $\beta$, and the clumping factor, $f_{cl}$. To lift this degeneracy more observational data from within the ALMA range is required.

\section{Acknowledgments}

The authors acknowledge support from the Science and Facilities Research Council (STFC). 

Computations were performed using the University of Birmingham's BlueBEAR HPS service, which was purchased through HEFCE SRIF-3 funds. See http://www.bear.bham.ac.uk.

The authors also thank the reviewer for their detailed comments.

\bsp

\label{lastpage}

\end{document}